# Smart patterned surfaces with programmable thermal emissivity and their design through combinatorial strategies

N. Athanasopoulos[a,1], N. J. Siakavellas[a]

[a]Department of Mechanical Engineering & Aeronautics, University of Patras, 26500, Patras, Greece

[1]Corresponding author. Tel.:+306946630065; fax: +302610997241. E-mail address: nathan@mech.upatras.gr (N. Athanasopoulos).

**Abstract**

The emissivity of common materials remains constant with temperature variations, and cannot drastically change. However, it is possible to design its entire behaviour as a function of temperature, and to significantly modify the thermal emissivity of a surface through the combination of different materials and patterns. Here, we show that smart patterned surfaces consisting of smaller structures (motifs) may be designed to respond uniquely through combinatorial design strategies by transforming themselves from 2D to 3D complex structures with a two-way shape memory effect. The smart surfaces can passively manipulate thermal radiation—without the use of controllers and power supplies—because their *modus operandi* has already been programmed and integrated into their intrinsic characteristics; the environment provides the energy required for their activation. Each motif emits thermal radiation in a certain manner, as it changes its geometry; however, the spatial distribution of these motifs causes them to interact with each other. Therefore, their combination and interaction determine the global behaviour of the surfaces, thus enabling their *a priori* design. The emissivity behaviour is not random; it is determined by two fundamental parameters, namely the combination of orientations in which the motifs open (n-fold rotational symmetry (rn)) and the combination of materials (colours) on the motifs; these generate functions which fully determine the dependency of the emissivity on the temperature.

Temperature control is one of the most common processes in Nature and man-made systems. By observing the manner in which plants and animals control their temperature (1–5) and their geometrical characteristics (6–8), we may deduce that Nature is a specialist in purely mechanistic thermal management strategies. Nature addresses thermal management issues—prevention of overheating and damages—through an efficient holistic and mechanistic design approach.

By utilising common paints (9) and by forming arrays in micro or in macroscale (10–14), certain emissivity values can be achieved that would allow the temperature regulation of a body and emittance direction. On the other hand, to achieve variable thermal emissivity properties, different approaches have been adopted according to the target application, namely the development of: **i)** advanced materials (15–17), **ii)** active metamaterials (18–20) through the formation of patterns in microscale, **iii)** active micro-electromechanical systems which incorporate materials of different thermo-optical properties (21,22), and **iv)** the development of systems, such as mechanical louvers (23) and morphing radiators, which conceal/reveal materials of different emissivity values using shape memory alloys (24,25). Furthermore, in architecture, large morph-able facades have been proposed for the shading of buildings (26–29). In this context, the present authors have preliminary studied the variability of thermal emissivity behaviour of a single self-shape structure (30).

Man-made and natural strategies combine the geometrical characteristics, the materials, and the patterns of the outer surface of a body to handle the thermal energy exchange in the nano-, micro- or macro-scale.

Based on the aforementioned mechanistic strategies, and taking them one step further, we theoretically and experimentally studied smart patterned surfaces† and identified their fundamental properties, which govern the global behaviour of the overall thermal energy emission. All necessary information was integrated to their material structure (shape transformation) and to the formed patterns by combining geometrical patterns and colour sequences. It is important to be mentioned that in the field of mechanical metamaterials (31–33) and thermal metamaterials for heat flux manipulation (34), interesting studies have revealed the importance of the manner in which the unit cells of metamaterials interact, as well as the impact of these unit cells on the local or global properties of the metamaterials.

## Generalised Approach on a Ditranslational Lattice

All surfaces can be represented as a ditranslational lattice, which is composed of unit cells (Fig. 1A and B). The lattice can be entirely tiled by motifs; each motif may have a different orientation on the lattice (Fig. 1C). The combination of motifs of different orientations constitutes a pattern (Fig. 1D).

The transformation of the motifs conceals (closed geometry) or reveals (open geometry) one of the materials (Fig. 2B and C), hereafter referred to as 'colours', and regulates the view factor of the patterned surfaces, thus enabling the realisation of a variable and programmable effective thermal emissivity ($\varepsilon_{eff}$) (Movies S2 and S3).

**Geometrical Transformation of the Motifs.** The folding of the structures may be realised through the use of shape memory materials only at certain temperatures (35,36). In this study, we developed anisotropic multilayer materials for which a temperature change generates internal stresses that cause the transformation of the motifs (Methods; Fig. S1, Movie S4), at any temperature level. These materials behave in a similar manner to the '4D-biomimetic materials (37–41). Owing to the large displacements of the multilayer material, non-linear phenomena appear (Methods) (42). The motifs are connected with non-deformable regions (detail in Fig. 1D); thus the assembly of the motifs, the overlaps, and the voids between them was avoided.

† In a broader context "patterned surfaces" may be referred to as "metasurfaces".

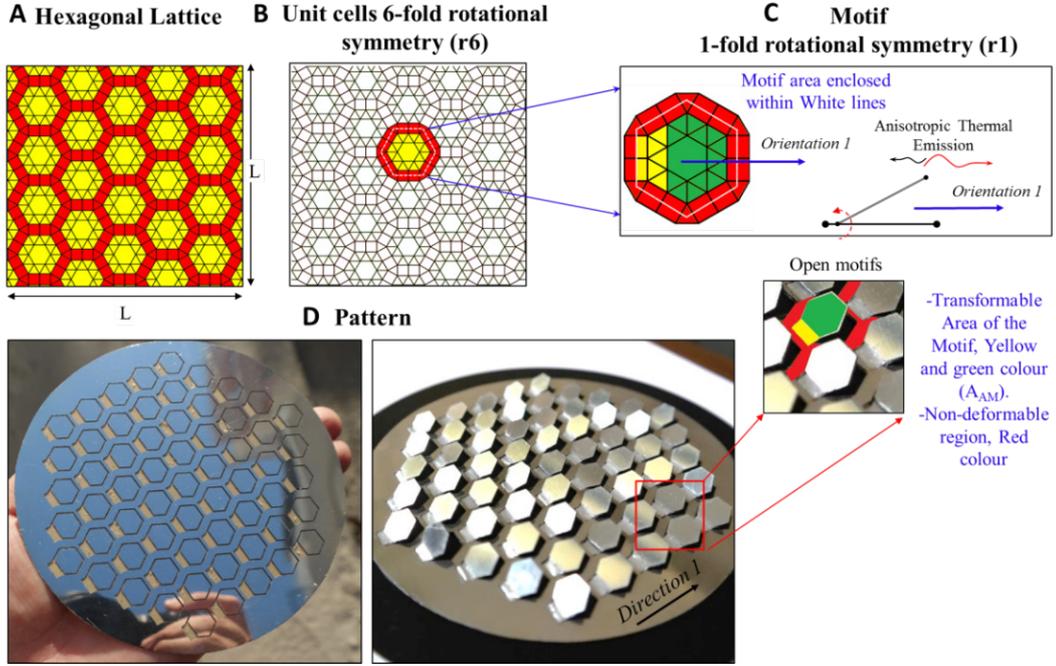

**Fig. 1.** Patterned surface incorporating motifs with 1-fold rotational symmetry (r1). (A) Di-translational hexagonal lattice. (B) 6-fold rotational symmetry unit cells. (C) Motif with 1-fold rotational symmetry. The possible orientations of the motifs on the di-translational lattice are related to the rotational symmetry of the motif and of the unit cells. (D) Developed patterned surface (Movie S1) and depiction of the transformable and non-deformable regions of the motifs.

**Fundamental parameters.** The transformation of the motifs is the 'driving force' in accomplishing our purpose; however, our scope is to investigate the impact of the interactions between the motifs on the thermal energy which is emitted from the patterned surfaces, for all directions and wavelengths (effective total hemispherical emissivity, $\varepsilon_{eff}(T)$) (Fig. 2D).

The behaviour of the generated emissivity curves is attributed mainly to: **i.** the combination (permutations) of colours (Fig. 2E) which have been applied onto the internal and external areas (positions/layers) of the motifs and to **ii.** the combination (permutations) of orientations in which the motifs open or close (n-fold rotational symmetry (rn)), (Fig. 1C and D). Therefore, we studied the interaction of two different ordered lists of elements with repetitions. Essentially, we focused on the order/sequence of the orientations of the motifs, the sequence of the applied colours, and their impact on the total hemispherical emissivity as a function of temperature of the patterned surface (global response).

The change in the emissivity is mainly due to the coexistence of different colours which have been placed on the internal and external surfaces of the motifs (Fig. 2E), and to the interaction thereof (successive reflections and absorptions). These materials have certain emissivity and absorptivity values; we assumed that they are grey ($\varepsilon_1 \neq \varepsilon_2$, $\alpha_1 \neq \alpha_2$, $a_1/\varepsilon_1 = a_2/\varepsilon_2 = 1$). Three different colours ($c_i$, where i = 1, 2, 3) may be combined and placed on three different motif layers/positions ($p$), $C[\{c_1, c_2, c_3\}, p]$. The simplest case is to select two different colours, and to set them to three different layers/positions $C[\{c_1, c_2\}, 3]$.

In addition, each motif may have a certain orientation in which it opens or closes; consequently, the motifs can anisotropically emit thermal radiation according to their rotational symmetry (1, 2, or n-fold rotational symmetry (rn)). A di-translational hexagonal lattice (43) may consist of unit cells of six possible different directions (Fig.1B). In this case, 1-fold rotational symmetry (r1´) motifs (Fig.1C) could open on a 6-fold rotational symmetry (r6) unit-cell, $P[\{1,2,3,...,6\},N] = \left(\dfrac{r6}{r1'}\right)^N$, where (P) is the number of possible permutations with repetitions (sequences), (N) is the number of motifs on the patterned surface, (rn) is the n-fold rotational symmetry of the unit cell, and (rn´) is the n´-fold rotational symmetry of the motif.

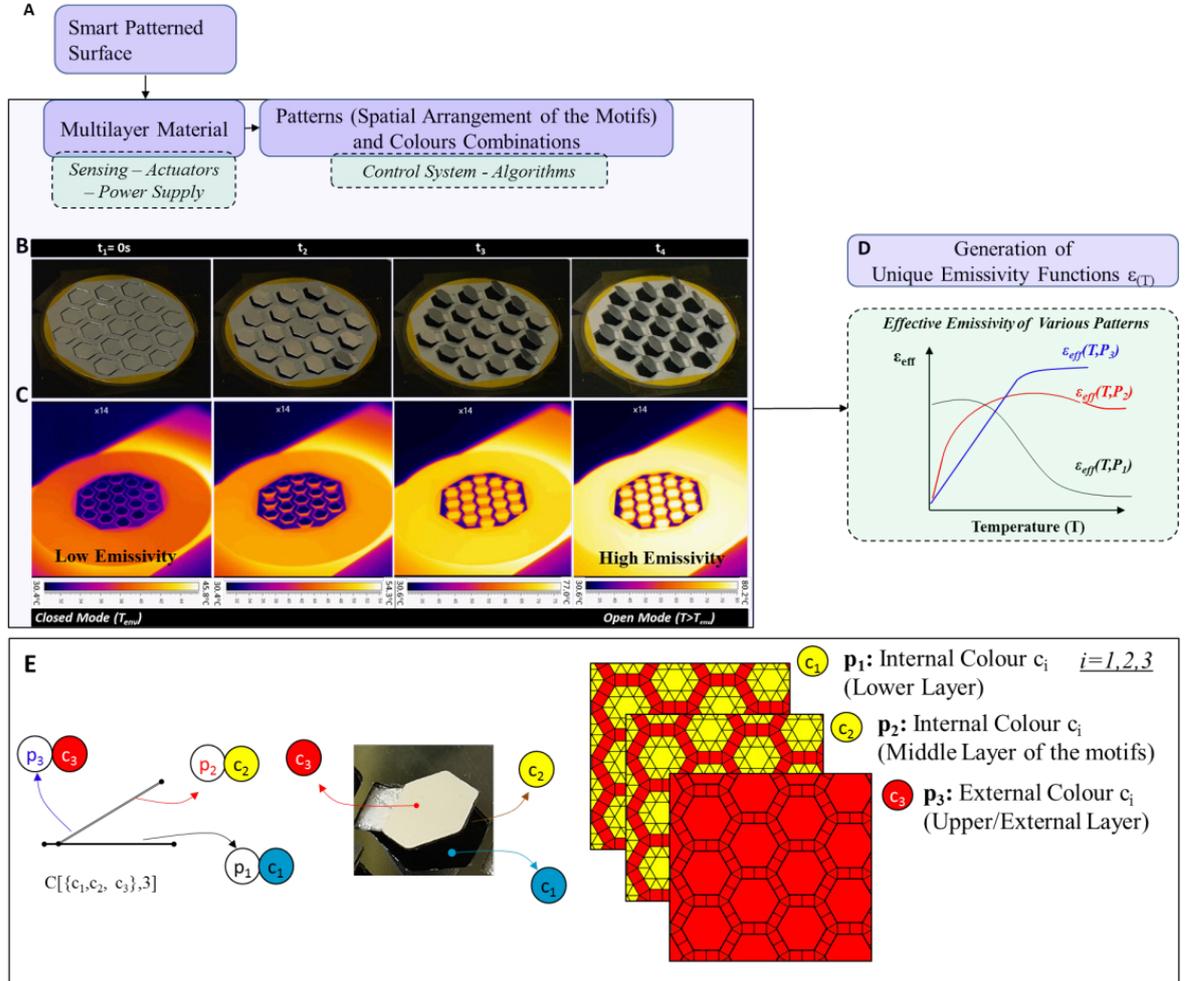

**Fig. 2.** Variable and programmable emissivity through smart patterned surfaces. (A) Flow diagram of the *modus operandi* of the materials. (B) Developed patterned surface on hexagonal di-translational lattice with variable emissivity as a function of temperature ($T_{env} \leq T \leq 80.2$ °C, $T_{max} \approx 100$ °C). The patterned surfaces can assume states from closed to open (Movie S2) or from open to closed (Movie S3). (C) Thermography during the heating stage. (D) Possible generated emissivity curves using different colour and orientation combinations. (E) Transformable region of the motif and different materials (colours, $c_i = 1,2,3$) with different emissivity values (ε) can be applied on three different layers (possible positions). Each motif could open facing a particular direction(s), concealing/revealing the internal layers to the environment.

The characteristics of the emissivity curve $\varepsilon_{eff}(T)$ can be controlled through the colour sequences (C) and the sequences of motifs (P) on a lattice (pattern). The sequence of the orientations of the motifs and the colours determines the maximum absolute change in emissivity ($\Delta\varepsilon_{max}$) for a specific temperature change ($\Delta T$) (global minimum and maximum of the curve), as well as the 'path' which the emissivity follows as the temperature of the body changes. Consequently, each sequence generates a function of $\varepsilon_{eff}(G_C^P,T)$ of certain characteristics, which can be manipulated through the orientation–colour coupling of these sequences. Here, (G) denotes the temperature depended generated function of ($\varepsilon_{eff}(T)$). The dominant behaviour of each $\varepsilon_{eff}(G_C^P,T)$ function is invariant to secondary parameters.

# Tiling a Strip (Monotranslational Lattice)

**Orientation Sequences.** By tiling a finite strip (mono-translational lattice) through the repetition of a motif along this strip, a variety of patterns are generated. The unit cell of the mono-translational lattice has a 2-fold rotational symmetry (r2) (Fig. 3). In this case, each motif with a 1-fold rotational symmetry (r1) may be oriented to face two different directions, such that the motif facing one direction is the mirror of the motif which faces the other direction (Fig. 3A and B). Therefore, in the strip problem, the motifs may receive only two discrete values, namely the right (R ≡ 1) and the left (L ≡ 2). For a combination of (N) motifs with 1-fold rotational symmetry, it becomes possible to generate P[{R,L},N] = [{1,2},N] = $2^N$ sequences. If (N) is an odd number, $2^N$ patterns can be generated; therefore, $2^N/2$ are the mirror patterns and $2^N/2$ are the unique generated patterns. Owing to the fact that the effective properties constitute a global property which is not related to the direction of the overall pattern—however, it is related to the orientation of the motifs—all mirror symmetries of the generated patterns are equivalent to each other for any colour sequence (C). If (N) is an even number, then $\left(2^N - 2^{N/2}\right)/2$ mirror patterns and $2^{N/2}$ self-similar (self-dual) mirror patterns can be generated. The self-similar patterns are the unique sequence for which the mirror of a pattern returns the exact same pattern with the exact same overall radiative behaviour (Fig. 3C). The unique generated patterns are equal to the total generated patterns minus the unique mirror patterns, $P = \left(2^N + 2^{N/2}\right)/2$.

**Colour Sequences.** Moreover, each motif may comprise a combination of colours. By applying two different colours, i.e. black (high emissivity = $\varepsilon_1 = c_1$) and silver (low emissivity = $\varepsilon_2 = c_2$), in three different available positions (p) on the layers of the motifs, the number of possible colour sequences is $C[\{c_1, c_2, ..., c_i\}, p] = c^p \xrightarrow[i=2]{p=3} C[\{c_1, c_2\}, 3] = 2^3$. The number of unique orientation and colour sequences and, consequently, the number of generated functions $\varepsilon_{\text{eff}}(G_C^P, T)$ of a finite-length strip can be expressed via the following equations.

$$G_C^N = G_{[\{c_1,c_2\},p]}^{[\{L,R\},N]} = C * P = c^p \frac{\left(2^N + 2^{N/2}\right)}{2}, \quad N = even$$

$$G_C^N = G_{[\{c_1,c_2\},p]}^{[\{L,R\},N]} = C * P = c^p \frac{\left(2^{N/2}\right)}{2}, \quad N = odd$$

(1)

We theoretically studied a plethora of different generated patterns; for some selected cases, we validated the results experimentally. Through the parametric numerical models we identified the fundamental properties that govern the overall emissivity behaviour and manipulate the emissivity curve (Methods). Assuming a grey diffuse body, ε(T) = α(T), we calculated the effective thermal emissivity through the following relation, which is the ratio of the total amount of energy that leaves the patterned surface to that emitted from a black-body area.

$$\varepsilon_{\text{eff}}(T) = \frac{Q}{A\sigma\left(T^4 - T_{env}^4\right)} \quad (2)$$

**1st level of classification – Change in emissivity.** Three levels of classification exist. The first-level the classification of the $\varepsilon_{\text{eff}}(G_C^P, T)$ functions concerns the different colour sequences; 1st group: increased emissivity $\Delta\varepsilon_{\text{eff}} > 0$, and 2nd group: decreased emissivity $\Delta\varepsilon_{\text{eff}} < 0$, as a function of temperature (Fig. 4). Figure 5A presents the generated functions of $\varepsilon(G_{[\{c_1,c_2\},3]}^{[1111]})$ for all colour sequences. Each colour sequence leads to a different discrete global max/min value for all unique patterns. By analysing the generated emissivity functions, we may observe that the colour sequences $\left[\{111\},\{121\},\{211\},\{221\}\right]$ generate emissivity functions with positive $\Delta\varepsilon$ ($\Delta\varepsilon > 0$) (1st group of Fig. 5A), whereas the inverse sequences $\xrightarrow{\overline{1}=2}_{\overline{2}=1}\left[\{222\},\{212\},\{122\},\{112\}\right]$ generate emissivity functions with a negative emissivity deviation ($\Delta\varepsilon < 0$) (2nd group of Fig. 5A). This is valid in the case where the motifs are transformed from the closed to the open state. In the case where the motifs transformed from the open to the closed state, the above statement is true only if we reverse the colour sequences.

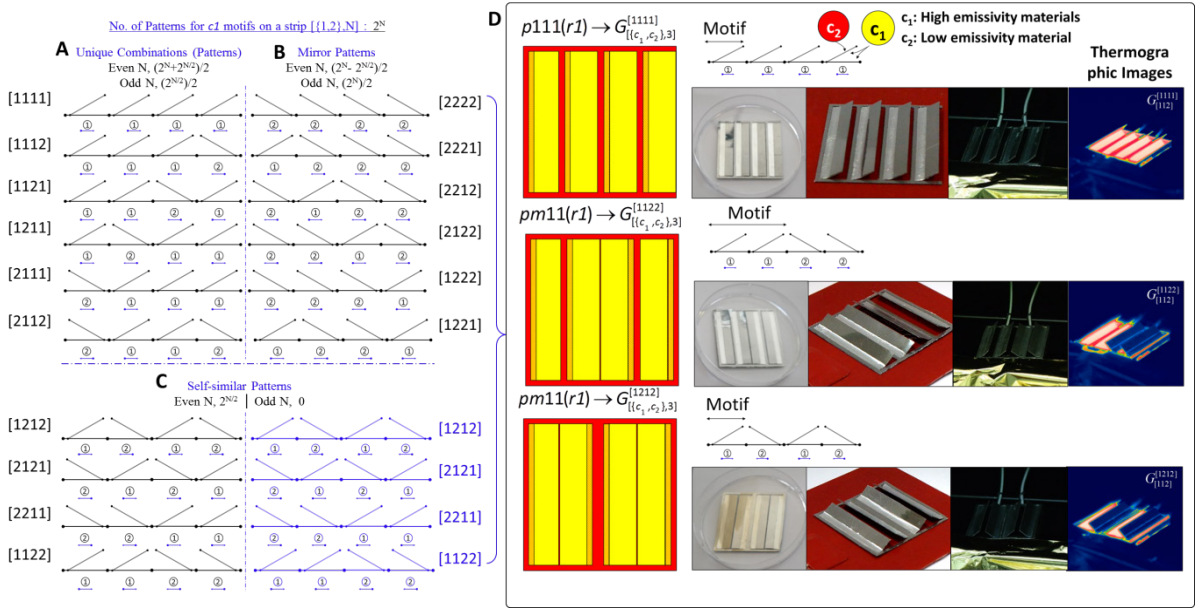

**Fig. 3.** Motifs on a strip. (A) Unique number of permutations of a pattern with N = 4. (B) Generated mirror patterns. (C) Generated self-similar patterns. (D) Selected patterns experimentally studied ($G_{[112]}^{[1111]}$, $G_{[112]}^{[1122]}$, $G_{[112]}^{[1212]}$) during different preliminary tests (Fig. S2, Movie S5).

**2nd level of classification – Linearity, convexity, and similarity.** The secondary-level classification is related to the orientation permutations/sequences of the motifs, and determines the linearity of the path that the curve will follow to reach the min/max values of ($\varepsilon_{eff}$); the subgroup (Fig. 5B) indicates the measure of linearity until the max/min value of the function is reached. Therefore, concave $\varepsilon(G_{[112]}^{[1111]})$, convex $\varepsilon(G_{[112]}^{[1212]})$, or 'linear' $\varepsilon(G_{[112]}^{[1122]})$ functions are generated (Fig. 5B) for all colour sequences $[\{c_1,c_2\},3]$. All generated functions of the subgroup are restricted within an upper limit $\varepsilon(G_{[112]}^{[1111]})$ and a lower limit $\varepsilon(G_{[112]}^{[1212]})$; the emissivity of all other patterns is restricted within these limits (Fig. S3B). The measure of linearity was correlated with the shape of the curve (44), and could be characterised and categorised using different mathematical quantities, such as the ellipticity, eccentricity or rectangularity. To compare the linearity of the generated functions of the subgroup, we employed ellipticity measures.

By measuring the ellipticity of a finite set of points, it is possible to classify our curves. The central moment $\mu_{pq}$ of the (pq) order is the following.

$$\mu_{pq} = \frac{1}{S}\sum\left(T - T_c\right)^p\left(\varepsilon - \varepsilon_c\right)^q \quad (3)$$

$$\left(\overline{T}, \overline{\varepsilon}\right) = \left(\frac{1}{S}\sum T_i, \frac{1}{S}\sum \varepsilon_i\right) \quad (4)$$

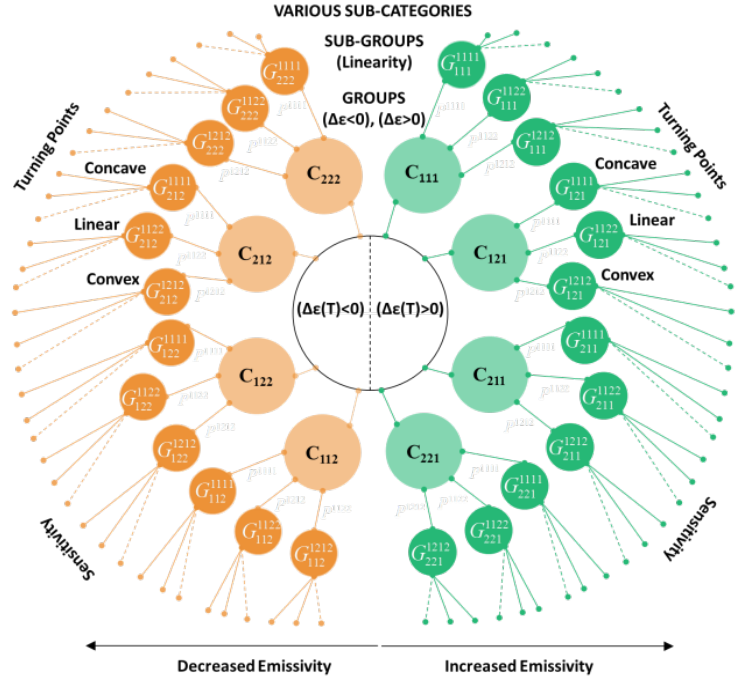

**Fig. 4.** Generation and classification of emissivity functions according to: i. Increased or decreased emissivity (2 major groups) for closed-to-open motifs, ii. Degree of linearity (8 sub-groups), and iii. Determination of sensitivity and turning points (sub-categories).

Here, $(\overline{T},\overline{\varepsilon})$ is the average value of each coordinate ($T_i$, $\varepsilon_i$). The linearity can be correlated with the ellipticity of the point of the generated function, and can be expressed as in the following.

$$a = \sqrt{2\left[\mu_{20} + \mu_{02} + \sqrt{(\mu_{20} - \mu_{02})^2 + 4\mu_{11}^2}\right]/\mu_{00}}$$

$$b = \sqrt{2\left[\mu_{20} + \mu_{02} - \sqrt{(\mu_{20} - \mu_{02})^2 + 4\mu_{11}^2}\right]/\mu_{00}} \quad (5)$$

$$\lambda = 1 - a/b, \quad 0 \leq \lambda \leq 1$$

The linear curves are denoted as $\lambda \approx 1$, whereas the non-linear yields values of $\lambda < 1$. The value of the linearity measure of the 1st curve $\varepsilon(G_{[112]}^{[1111]},T)$ is $\lambda = 0.893$, of the 2nd curve $\varepsilon(G_{[112]}^{[1122]},T)$ is $\lambda = 0.97$, and of the 3rd $\varepsilon(G_{[112]}^{[1212]},T)$ is $\lambda = 0.91$. Pattern $G_{[112]}^{[1122]}$ generates functions which are almost linear (Fig. 5B) prior to the function reaching a plateau.

Other simple tools are related to the identification of the convexity of the generated function. The convexity of the discrete function $\varepsilon(G_C^P,T)$ between three points can be expressed using the following simple inequality.

$$\varepsilon(i-1) + \varepsilon(i+1) - 2\varepsilon(i) > 0 \quad (6)$$

**3rd level of classification – Sensitivity.** Other parameters which affect the overall radiative behaviour of the patterned surface are related to i) the physical properties of the materials, ii) the shape and iii) the distance between the motifs, and iv) thickness of the multilayer material. In the case where the thickness of the multilayer material decreases, the emissivity function maintains its behaviour; however, with a greater sensitivity (Fig. 5C). It was found that all generated functions adapt their form to achieve the min/max values within a smaller temperature span; however, their dimensionless form has extremely similar characteristics. The value of the linearity measure for the two different thicknesses ($H_1$, $H_2$) of the 1st curve $\varepsilon(G_{[112]}^{[1111]}, T)$ ranges within $0.893 < \lambda < 0.930$, of the 2nd curve $\varepsilon(G_{[112]}^{[1122]}, T)$ ranges within $0.97 < \lambda < 0.99$, and of the 3rd $\varepsilon(G_{[112]}^{[1212]}, T)$ ranges within $0.91 < \lambda < 0.930$ (Table S2).

Moreover, we used simple schemes to validate the similarity between the shapes of the curves for the two different thicknesses, such as the normalised correlation and Pearson's correlation coefficient. Their values indicated the similarity of the two generated curves with different thickness for the same colour and orientation sequences. The measure of similarity receives values between $-1 \leq s \leq 1$; for $s \approx 1$, the correlation/similarity is excellent.

$$s_1 = \frac{\left( i \sum \varepsilon_C^P(i)_1 \varepsilon_C^P(i)_2 - \sum \varepsilon_C^P(i)_1 \sum \varepsilon_C^P(i)_2 \right)}{\sqrt{i \sum \left( \varepsilon_C^P(i)_1 \right)^2 - \left( \sum \left( \varepsilon_C^P(i)_1 \right) \right)^2} \sqrt{i \sum \left( \varepsilon_C^P(i)_2 \right)^2 - \left( \sum \left( \varepsilon_C^P(i)_2 \right) \right)^2}} \quad (7)$$

$$s_2 = \frac{\sum \varepsilon_C^P(i)_1 \varepsilon_C^P(i)_2}{\sqrt{\sum \left( \varepsilon_C^P(i)_1 \right)^2 \sum \left( \varepsilon_C^P(i)_2 \right)^2}} \quad (8)$$

Here, $\varepsilon_1(i)$ and $\varepsilon_2(i)$ represent two emissivity curves which have been generated by the smart patterns of different thickness. The similarity between the two emissivity curves for three different patterns are: $s = 0.996$ for $\varepsilon(G_{[112]}^{[1111]}, T)$, $s = 0.998$ for $\varepsilon(G_{[112]}^{[1122]}, T)$, and $s = 0.999$ for $\varepsilon(G_{[112]}^{[1212]}, T)$ (Table S2). The emissivity curves maintain their characteristics; they only became 'distorted', and were located within a smaller $\Delta T$ (Fig. 5C). By calculating the correlation coefficients of three different functions—$\varepsilon(G_{112}^{1212})$, $\varepsilon(G_{112}^{1122})$, $\varepsilon(G_{112}^{1111})$—versus the thickness, we concluded that the similarity in all cases ranged within 99.6–99.9%. Therefore, we may deduce that only two parameters determine the main characteristics of the curve, namely the orientation patterns and the colour sequences.

In a similar manner, if the motif has a fully deformable region, the behaviour remains the same; this region affects only the sensitivity of the curve ($\Delta\varepsilon_{max}/\Delta T$) and may slightly modify the global min/max values (Fig. S3D). Owing to the finite dimensions (length) of the pattern, the emissivity concerns only the projection of its geometry and not the entire exposed area of the pattern. This causes uncertainties owing to the edge effects of the pattern. Thus, we developed models with a different number of motifs, $4 < N < 65$, as a function of temperature. Only small deviations were present because the edge effects were nullified (Fig. S3E and S3F).

**Approximating a predetermined curve.** Moreover, we selected a convex bounded equation $\varepsilon(T)$ of certain characteristics with the purpose of predicting the sequences of colours and orientations that would yield the selected equation. We used a predetermined bounded equation $\varepsilon(T)$ of the form $\varepsilon(T) = \dfrac{a_1}{b_1 + b_2 \left( e^{-cT^d} \right)} + a_2$, which can be configured to produce linear, concave or convex paths. Power (d) regulates the linearity of the equation; $a_1$, $a_2$, $b_1$, and $b_2$ regulate $\Delta\varepsilon$ and $\Delta T$. By setting the lower and upper limits to $\varepsilon_{min}= 0.1$ and $\varepsilon_{max}= 0.6$, respectively, within the temperature change of $\Delta T = 80$ °C, the following equation was obtained: $\varepsilon(T) = \dfrac{2}{1+0.333\left(e^{-0.1T}\right)} + 1.4$. After examining the sequences of colours and orientations, we concluded that $\varepsilon(G_{[122]}^{[1111]},T)$ best approximates the aforementioned equation. These specific sequence produced a curve whose shape was 99.89% (Eq. 7) similar to that of the predetermined equation, and with the same measure of linearity ($\lambda_{pred}$ = 0.86, $\lambda_{gener}$ = 0.85) (Eq. 5) and emissivity change ($\Delta\varepsilon_{pred}$ = 0.5, $\Delta\varepsilon_{gener}$ = 0.517) (Fig. 5D).

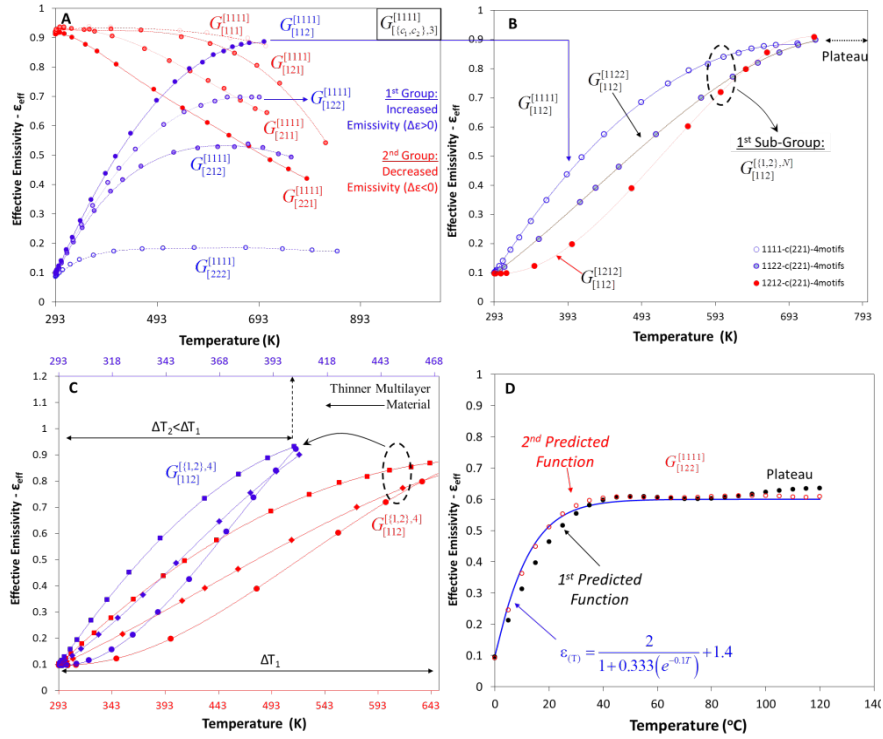

**Fig. 5.** Generated effective emissivity for the curves of the finite-strip problem according to the 1st and 2nd classification. (A) Two generated groups of functions for pattern $G_C^{[1111]}$ (1st classification). (B) Generated subgroups of functions (2nd classification) for combinations $G_{[112]}^{[1111]}, G_{[112]}^{[1122]}, G_{[112]}^{[1212]}$. (C) Transformation of the functions for a more sensitive multilayer film. (D) Predicted generated emissivity curve $\varepsilon\left(G_{[122]}^{[1111]}, T\right)$ vs the predetermined curve.

**Experimental Verification.** Finally, we experimentally verified the theoretical results through a plethora of measurements, using calorimetric techniques (SI2). We selected three particular patterns $\varepsilon(G_{[112]}^{[1111]})$, $\varepsilon(G_{[112]}^{[1122]})$, $\varepsilon(G_{[112]}^{[1212]})$ because they have different measures of linearity, and they define the extreme lower and higher limits of the possible generated thermal emissivity functions (Fig. 6). The change in emissivity is $\Delta\varepsilon \approx 0.47$ for all curves. The experimental results generated three curves of different measures of linearity, which is in agreement with our prediction within the temperature span of $\Delta T \approx 80$ °C. Small differentiations in the behaviour of the acquired and the theoretical curves may be present owing to the specular nature of the aluminium surface, as well as the inequality of the absorptivity and the emissivity of the external aluminium surface, $\alpha_{al} > \varepsilon_{al}$. The experimentally studied patterned surfaces during the heating and cooling stage are presented in Fig. 3D, Fig. S2 & Movie S5.

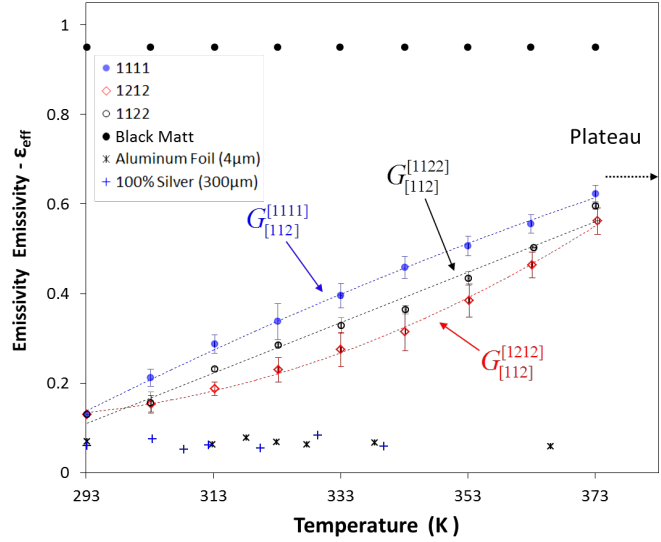

**Fig. 6.** Experimentally generated effective emissivity functions of the finite-strip problem according to the 1$^{st}$ for the above mentioned patterns; $G_{[112]}^{[1111]}, G_{[112]}^{[1122]}, G_{[112]}^{[1212]}$.

## Conclusions

By identifying and handling these fundamental properties—orientation and colour sequences—we designed the thermal emissivity function of a patterned surface. The classification of the generated curves and the similarities owing to the existence of invariant properties limit the number of combinations that need to be considered.

We drastically altered its value at different temperature levels; thus, we developed integrated, low-weight, cost-effective, and programmable thermal-management materials.

The present work can significantly contribute to the future thermal design of various energy systems— such as buildings and satellites/spacecraft for Space exploration—and sensors for the directional identification of a heat source or for handling different wavelengths. Furthermore, this work may lead to the development of "4D-5D materials" and thermal adaptable materials.

## Materials and Methods

**Material Structure.** We can tile a surface using a combination of motifs, which form a pattern. The patterned surfaces have an overall area of (A = N×A$_M$), where A$_M$ is the region of the motif (depicted as red, green, and yellow). The motif consists of a non-deformable region (depicted as red), the deformable region A$_{AM}$ (depicted as yellow and green) and a region which may be either deformable or non-deformable (depicted as green), (Fig. S1). The deformable are may be partially or fully deformable. The higher the fraction $F_M = A_{AM}/A_M \leq 1$ and the ratio $\varepsilon_1/\varepsilon_2$ are—which correspond to the inner and outer surfaces, respectively—the higher the change of the emissivity ($\Delta\varepsilon_{max}$).

The deformable regions are very responsive to temperature, presenting extremely large deformations (Fig. S1F; Movie S4). These materials are similar to the "4D-biomimetic materials", which become activated in proportion to the stimulus (humidity (37,39-40) or temperature) and can be manufactured via low-cost techniques. The mismatch of the coefficient of thermal expansion (CTE) between the anisotropic layers creates materials that are very sensitive in temperature and alter their shape drastically owing to the developed internal stresses and their anisotropic nature. The sequence of the layers and the materials of the non-deformable and deformable regions that were used for the development of various patterned surfaces are presented in Fig. S1 & Table S1.

**Theoretical considerations of the combinatorial strategy and parametric numerical modelling.**

The generated discrete functions $\varepsilon\left(G_C^P, T\right) = \varepsilon\left(G_{[\{c_1,c_2\},p]}^{[\{R,L\},N]}, T\right) \xrightarrow[c_1 \equiv 1, c_2 \equiv 2]{R \equiv 1, L \equiv 2} \varepsilon\left(G_{[\{1,2\},p]}^{[\{1,2\},N]}, T\right)$ of a mono-translational or di-translational lattice can be expressed as the combination of two finite ordered lists of elements. Regarding the strip problem, the combination of two finite ordered lists of elements are P[{R,L},N] and C[{$c_1,c_2$},p]. A motif with 1-fold rotational symmetry-r1 on a unit cell with a 2-fold rotational symmetry-r2 generates two different orientations {R,L}. For N = 4 motifs, the generated sequences are the following: [{RRRR}, {RRRL}, {RLRR}, {LRRR}, {RLRL},…,{LLLL}], (Table S3). In a similar manner, the colour sequences for two colours in three positions (p) are [{$c_1,c_1,c_1$}, {$c_1,c_1,c_2$}, {$c_1,c_2,c_1$},...,{$c_2,c_2,c_2$}]. All mirror patterns lead to equivalent emissivity values e.g. the sequence $\left\{[RLRR] \equiv [LRLL] \xrightarrow[L \equiv 2]{R \equiv 1} [1211] \equiv [2122]\right\}$, and generate exactly the same emissivity function $\varepsilon\left(G_C^{[1211]}, T\right) = \varepsilon\left(G_C^{[2122]}, T\right)$. A first screening excludes all mirror [1211] ≡ [2122] or self-similar patterns [1122] ≡ [1122], thus reducing the number of patterns which lead to unique solutions. As previously described, two different cases are presented in the mono-translational lattice: N = odd number and N = even number. If (N) is an odd number, $2^N$ patterns can be generated; therefore, $2^N/2$ are the mirror patterns and $2^N/2$ are the unique generated patterns.

For N = even number, the generated mirror patterns are $\neq 2^N/2$, and the unique generated patterns are $\neq 2^N/2$. This is valid owing to the existence of the self-similar patterns ($2^{N/2}$). As a consequence, the unique generated patterns are equal to the total generated patterns minus the unique mirror patterns, $P = 2^N - \left(2^N - 2^{N/2}\right)/2 = \left(2^N + 2^{N/2}\right)/2$, or to the sum of the unique generated patterns and the self-similar mirror patterns, $P = \left(2^N - 2^{N/2}\right)/2 + 2^{N/2} = \left(2^N + 2^{N/2}\right)/2$.

In the case of one colour, all the sequences generate approximately the same curve. In contrast, the combination of at least two different colours generates an entire family of different emissivity curves (Fig. 4 and Fig. S4). A strip with 4 motifs (positions) has 10 unique patterns, and the use of 2 different colours in 3 possible positions generates 80 unique sequences. In the Table S4, we list all unique generated functions for N ≤ 7 for 1 colour, 2 colours, and 3 colours. It is important to be mentioned that the exact integer sequences of the unique mirror patterns and the unique number of the generated patterns $\forall N$ may be found in the On-Line Encyclopedia of Integer Sequences (A007179 & A051437) in two studies of other scientific disciplines (45,46).

All possible unique sequences for N=4 motifs were solved through steady-state or transient coupled thermo-mechanical models for different temperature levels using the COMSOL Multiphysics. The physical properties, as well as the dimensions, are presented in detail (SI2). These models generate all the unique discrete emissivity functions versus temperature. In the case of large displacements, non-linear phenomena appear. The non-linear behaviour originates from the geometrical non-linearity that is due to the small thickness and the large displacements and rotations of the multilayer material, and not from the properties of the material itself (42). For this reason, the strain tensor with the non-linear terms should be considered. In our case, it is necessary to solve a coupled thermo-mechanical and geometrically non-linear problem which incorporates the interactions of the body caused by thermal radiation as well. The Green–Lagrange strain tensor represents the strains, and the Second Piola–Kirchoff stress tensor represents the stresses. We used structured quadrilateral elements to model the overall phenomenon.

$$e_{el} = \frac{1}{2}\left[(\nabla u)^T + \nabla u + (\nabla u)^T \nabla u\right] - a(T - T_{ref}) \quad (9)$$

Owing to the complexity of the problem, we developed patterned surfaces which can be modelled through 2D plain strain problems (Fig. 3). We solved these models parametrically in order to study the interactions between the motifs and to predict their final geometry, the temperature field, and the effective emissivity of the patterned surfaces. All the unique orientation and colour sequences were solved and classified accordingly (Fig. 4). The parameters of interest are the radiosity, emissivity, temperature, and the irradiation. The radiosity leaving a surface is defined by

$$J_i = \rho_i G_i + \varepsilon_i \sigma T_i^4 \quad (10)$$

Here, ($\rho_i$) and ($\varepsilon_i$) are the surface reflectivity and emissivity, respectively, and (i) denotes the surface of the low- or the high-emissivity material. In general, the irradiation, (G), of the surface can be written as a sum, as in the following.

$$G_i = G_{m,i} + F_{a,i} \sigma T_a^4 \quad (11)$$

Here, ($G_m$) is the mutual irradiation from other boundaries, ($F_a$) is an ambient view factor, and ($T_a$) is the assumed far-away temperature in the directions included in ($F_a$). In fact, ($G_m$) is the integral over all visible points of a differential view factor (F) multiplied by the radiosity (J) of the corresponding source point. In the discrete model, ($G_m$) may be expressed as the product of a view factor matrix and a radiosity vector.

$$J_i = \rho_i \left[G_{m,i}(J) + F_{a,i} \sigma T_a^4\right] + \varepsilon_i \sigma T_i^4 \quad (12)$$

Assuming an ideal grey body, Eq. 12 becomes

$$J_i = (1 - \varepsilon_i)\left[G_{m,i}(J) + F_{a,i} \sigma T_a^4\right] + \varepsilon_i \sigma T_i^4, \quad i = 1, 2, \ldots, k. \quad (13)$$

Equation (12) results in an equation system in (*J*), which is solved in parallel with the temperature equation, (*T*); (k) expresses the number of surfaces which the overall structure consists of. We predicted the geometric transformation of the motifs and the change in the ambient view factor during their interaction for two different cases—partially and fully deformable area—by solving the transient problem (Movies S6 and S7).

**Measurements and apparatuses.** We used a comparative calorimetric method under vacuum to measure the effective emissivity (SI2).

**Other measured and developed smart materials.** We investigated a patterned surface which incorporated a ditranslational pattern with a rectangular lattice pattern (4 × 4 motifs). The emissivity increased within the temperature range of 20 °C to 57 °C. For higher temperatures, the emissivity remained almost constant. The developed surfaces could changed their emissivity Δε = 0.38 within a very small temperature deviation (ΔT = 37 °C), (Fig. S5).

1. Willmer P, Stone G, Johnston I (2005) *Environmental physiology of animals* doi:10.1007/s13398-014-0173-7.2.

**SI** has been included in this research article.

**Acknowledgements** This research is implemented through IKY scholarships programme and co-financed by the European Union (European Social Fund - ESF) and Greek national funds through the action entitled "Reinforcement of Postdoctoral Researchers" of the National Strategic Reference Framework (NSRF) 2014 – 2020.

We would like to thank Ms Polyxeni Souridi for the final editing of the English manuscript and Dr Roberto Guzman de Villoria for the discussions.

**Author Contributions** The manuscript was written by both authors. The central idea and the problem were conceived and formulated by N.A. N.J.S and N.A. formulated and developed the numerical models, analyse the results and conduct the experiments. The materials were developed by N. A.

# SUPPLEMENTARY FIGURES AND TABLES - SI1

# Smart patterned surfaces with programmable thermal emissivity and their design through combinatorial strategies

N. Athanasopoulos[a,1], N. J. Siakavellas[a]

[a]Department of Mechanical Engineering & Aeronautics, University of Patras, 26500, Patras, Greece

[1]Corresponding author. Tel.:+306946630065; fax:+302610997241. E-mail address: nathan@mech.upatras.gr (N. Athanasopoulos).

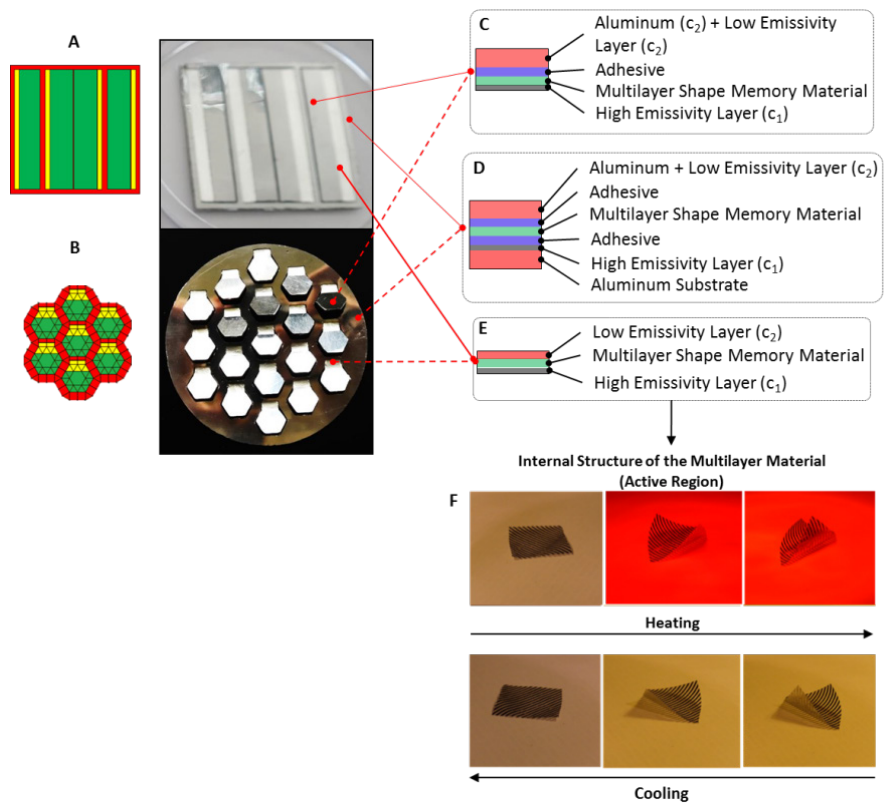

**Fig. S1.** Structure of the smart patterned surfaces. (A,B) Smart patterned surfaces with variable emissivity. (C-E) Deformable and non-deformable multilayer structure. (F) Multilayer shape memory material (Movie S4).

**Table S1.** Internal structure of the multilayer patterned surfaces at different regions.

|   | Deformable Region (Yellow) | Non-deformable Region (Red) | Deformable or Non-deformable Region (Green) |
|---|---|---|---|
| 1. | Oriented PE | Aluminium substrate | Oriented PE |
| 2. | Adhesive | Graphite coating | Adhesive |
| 3. | Aluminium strips | Oriented PE | Aluminium strips |
| 4. | Aluminium film (5 μm) | Adhesive | Aluminium film |
| 5. |  | Aluminium strips (%) | Adhesive |
| 6. |  | Aluminium film | Polished Aluminium |
| 7. |  | Adhesive |  |
| 8. |  | Polished Aluminium |  |

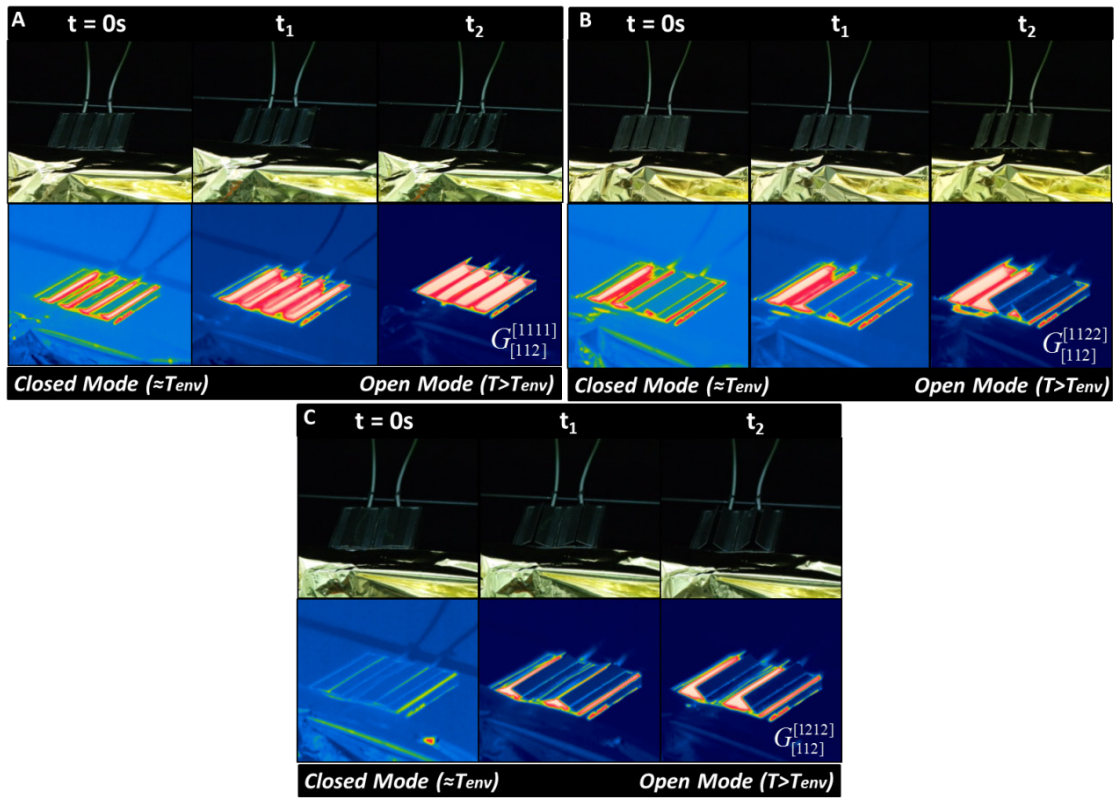

**Fig. S2.** The developed studied surfaces, incorporating rectangular motifs on a strip with dimensions of (54 × 54 mm), and thermographic images. (A) Pattern $G^{[1111]}_{[112]}$ (p111). (B) Pattern $G^{[1122]}_{[112]}$ (pm11). (C) Pattern $G^{[1212]}_{[112]}$ (pm11) (Movie S5).

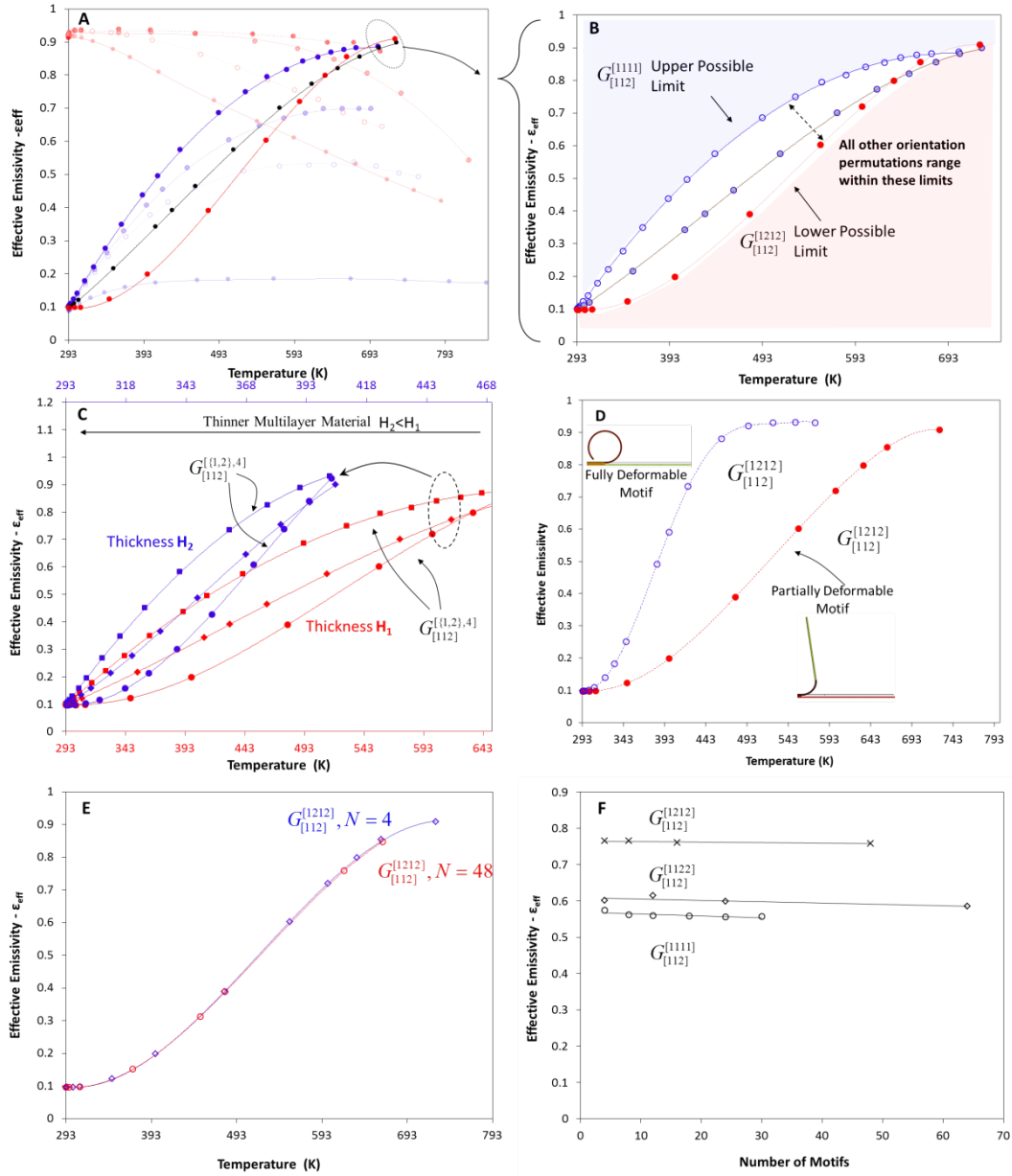

**Fig. S3.** Classification according to the behaviour (response) of the patterned surfaces. (A) Generated groups for P[1111] and for all colour sequences, $G^{[1111]}_{[\{c_1,c_2\},3]}$. (B) All orientation sequences P[{1,2},N] generate functions of different linearity, whereas the extreme sequences $G^{[1111]}_{[\{c_1,c_2\},3]}$, $G^{[1212]}_{[\{c_1,c_2\},3]}$ form upper and lower limit. (C) Generation of similar solutions as a function of the thickness of the multilayer material (invariant behaviour). (D) Comparison between the generated functions for a partial and a fully deformed motif for pattern $G^{[1212]}_{[\{c_1,c_2\},3]}$. (E) Generated function $\varepsilon\left(G^{[1212]}_{[\{c_1,c_2\},3]}, T\right)$ for N = 4, N = 48 motifs as a function of temperature. (F) Emissivity as a function of the number of motifs for three different orientation permutations.

**Table S2.** Linearity and similarity measures. Linearity of the curves for different multilayer thickness and similarity between these curves.

|  |  | $\varepsilon(G^{[1111]}_{[112]}, T)$ | | $\varepsilon(G^{[1122]}_{[112]}, T)$ | | $\varepsilon(G^{[1212]}_{[112]}, T)$ | |
| --- | --- | --- | --- | --- | --- | --- | --- |
|  |  | $H_1$ | $H_2$ | $H_1$ | $H_2$ | $H_1$ | $H_2$ |
| **Convexity** |  | Concave Nonlinear < 1 | | Linear ≈ 1 | | Convex Non-linear < 1 | |
| **Linearity** $0 \leq \lambda \leq 1$ | $\lambda_1$ | 0.893 | 0.930 | 0.970 | 0.99 | 0.930 | 0.910 |
| **Similarity** $-1 \leq s \leq 1$ | $s_1$ | 0.996 | | 0.998 | | 0.999 | |
|  | $s_2$ | 0.999 | | 0.999 | | 0.999 | |

**Table S3.** Generated Sequences as a Function of the Number of Motifs on a Strip. Comparison between the generated unique and mirror patterns for odd and even N.

| | | N = 1 | | N = 2 | | N = 3 | | N = 4 | |
|---|---|---|---|---|---|---|---|---|---|
| | | Pattern | Mirror Pattern | Pattern | Mirror Pattern | Pattern | Mirror Pattern | Pattern | Mirror Pattern |
| Sequences | 1 | R | L | RR | LL | RRR | LLL | RRRR | LLLL |
| | 2 | | | RL | RL | RRL | RLL | RRRL | RLLL |
| | 3 | | | LR | LR | RLR | LRL | RLRL | RLRL |
| | 4 | | | | | LRR | LLR | LRRR | LLLR |
| | 5 | | | | | | | LRRL | RLLR |
| | 6 | | | | | | | LRLR | LRLR |
| | 7 | | | | | | | RRLR | LRLL |
| | 8 | | | | | | | RLRR | LLRL |
| | 9 | | | | | | | LLRR | LLRR |
| | 10 | | | | | | | RRLL | RRLL |

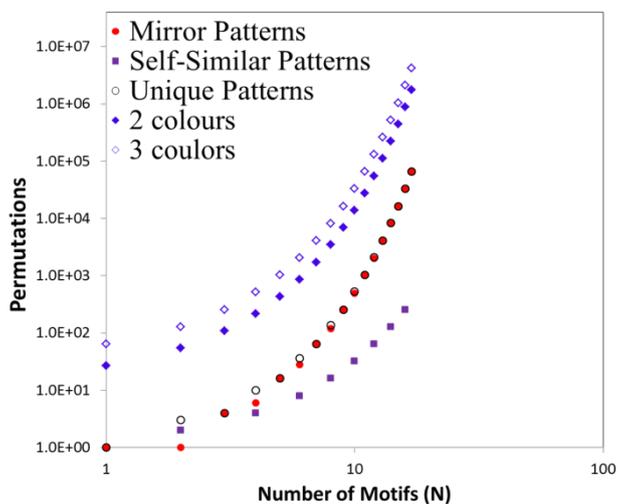

**Fig. S4.** Generated combinations of unique colour–orientation patterns as a function of the number of motifs on a finite strip.

**Table S4.** Possible Rotational (orientation) and Colour Permutations. Calculation of the unique patterns and functions for N = 1 to N = 7 motifs with 1-fold rotational symmetry, 1–3 colours, and p = 3 positions on the motifs.

| Motifs (N) | Possible Patterns [{L,R},N] | Mirror Patterns ($P_M$) | Self-similar Patterns ($P_S$) | Unique Sequences (P) | Unique Functions for $G^{[\{1,2\},N]}_{[\{c_1\},p]}$ | Unique Functions for $G^{[\{1,2\},N]}_{[\{c_1,c_2\},p]}$ | Unique Functions For $G^{[\{1,2\},N]}_{[\{c_1,c_2,c_3\},p]}$ |
|---|---|---|---|---|---|---|---|
| Even | $2^N$ | $(2^N-2^{N/2})/2$ | $2^{N/2}$ | $(2^N+2^{N/2})/2$ | $c^p(2^N+2^{N/2})/2$ | | |
| Odd | | $2^N/2$ | 0 | $2^N/2$ | $c^p(2^N)/2$ | | |
| 1 | 2 | 1 | 0 | 1 | 1 | 8 | 64 |
| 2 | 4 | 1 | 2 | 3 | 3 | 24 | 128 |
| 3 | 8 | 4 | 0 | 4 | 4 | 32 | 256 |
| 4 | 16 | 6 | 4 | 10 | 10 | 80 | 512 |
| 5 | 32 | 16 | 0 | 16 | 16 | 128 | 1024 |
| 6 | 64 | 28 | 8 | 36 | 36 | 288 | 2048 |
| 7 | 128 | 64 | 0 | 64 | 64 | 512 | 4096 |

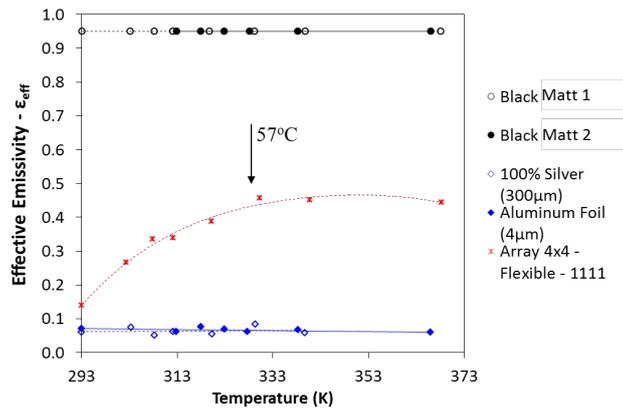

**Fig. S5.** Developed patterned surface (p1(r1)) with C[{1,2},3], N = 4 × 4 motifs. Total hemispherical emissivity as function of temperature.

# Smart patterned surfaces with programmable thermal emissivity and their design through combinatorial strategies

N. Athanasopoulos[a,1], N. J. Siakavellas[a]

[a]Department of Mechanical Engineering & Aeronautics, University of Patras, 26500, Patras, Greece

[1]Corresponding author. Tel.:+306946630065; fax:+302610997241. E-mail address: nathan@mech.upatras.gr (N. Athanasopoulos).

*The following information is provided for the replication of numerical models and experimental results.*

## I. Supplementary Methods

**1. Numerical modelling details:** COMSOL Multiphysics was used for the modelling of the coupled problem. All parametric models were executed on a powerful workstation, namely the Fujitsu CELSIUS R930 (two parallel processors, 24 cores) with two Intel Xeon E5-2697 v2 processors (2.70 GHz, 30 MB cache) Turbo Boost (256 GB RAM). Owing to the complexity of the problem, we developed patterned surfaces which can be modelled through 2D plain strain problems (Supplementary Movies 6 and 7). The 3D models may use more than 512 GB of RAM for a smart surface with 36 motifs because of the thermal radiation coupling. By solving the parametric models, we calculated the temperature field of the surface and the total hemispherical emissivity. All material properties and model dimensions are listed in Tables S5 and S6.

**Table S5.** Material Properties.

| Property | Symbol | Value | Units |
|---|---|---|---|
| Emissivity, $c_1$ | $\varepsilon_1$ | 0.95 | - |
| Emissivity, $c_2$ | $\varepsilon_2$ | 0.075 | - |
| Thermal conductivity | $k_1$ | 155 | W/mK |
| Thermal conductivity | $k_2$ | 0.4 | W/mK |
| Heat capacity | $cp_1$ | 893 | J/kg·K |
| Heat capacity | $cp_2$ | 2000 | J/kg·K |
| Young Modulus | $E_1$ | 69 | GPa |
| Young Modulus | $E_2$ | 0.5 | GPa |
| Poisson | $v_1$ | 0.33 | - |
| Poisson | $v_2$ | 0.4 | - |
| Coefficient of thermal expansion | $a_1$ | 200 | $(10^{-6})$ m/m°C |
| Coefficient of thermal expansion | $a_2$ | 23.2 | $(10^{-6})$ m/m°C |

**Table S6.** Dimensions of the motifs and the patterns for the numerical modelling, for partially and fully deformable active regions.

| | | Thickness = 125 μm Motifs, N = 4 | | Thickness = 25 μm Motifs, N = 4 | |
|---|---|---|---|---|---|
| | | Case_1 (Partially deformable) | Case_2 (Fully deformable) | Case_1 (Partially deformable) | Case_2 (Fully deformable) |
| l₁ (Length of the motif) | mm | 8.5 | 8.5 | 8.5 | 8.5 |
| μ₁ (Length of the non-deformable area) | | 6 | 0 | 6 | 0 |
| μ₂ (Length of the deformable area) | | 2 | 8 | 2 | 8 |
| μ₃ (Length of the non-deformable area) | | 0.5 | 0.5 | 0.5 | 0.5 |
| L (Total length of the patterned surface) | | 34 | 34 | 34 | 34 |
| A_A (Fraction of the active to the non-deformable area) | - | 0.25 | 1 | 0.25 | 1 |
| A_AM (Fraction of non-deformable area to the total area) | - | 0.94 | 0.94 | 0.94 | 0.94 |
| h₁ (Thickness of the 1st layer) | μm | 100 | 100 | 20 | 20 |
| h₂ (Thickness of the 2nd layer) | μm | 25 | 25 | 5 | 5 |

To ensure that the models accurately predict the deformed geometry, several models were developed and compared with results from analytical solutions. The generalized solution for the analytical prediction of any multilayer material with rectangular geometry was expressed as a system of equations (47). The trimaterial which we used can be modelled using a simplified equation for the bilayer material owing to the very small thickness of the middle adhesive layer. The curvature of the solved models was compared with that of the analytical solutions for the bilayer material. The curvature of a bilayer material can be calculated as:

$$\kappa = \Delta T (a_1 - a_2) \frac{6 E_1 E_2 h_1 h_2 (h_1 + h_2)}{E_1^2 h_1^4 + 4 E_1 E_2 h_1^3 h_2 + 6 E_1 E_2 h_1^2 h_2^2 + 4 E_1 E_2 h_1 h_2^3 + E_2^2 h_2^4}, \quad (S1)$$

where $\kappa$ is the curvature, $(\alpha_1, \alpha_2)$ are the coefficients of thermal expansion, $(E_1, E_2)$ are the moduli of elasticity, $(h_1, h_2)$ are the layer thicknesses, and $\Delta T$ is the temperature difference. We compared the curvature of the geometrically non-linear models with that of the analytical solutions for different temperature levels. The results are listed in Table S7.

**Table S7.** Comparison of numerical and analytical results of the calculated curvature for the non-linear thermo-mechanical problem at different temperature levels.

| Temperature change (°C) | Radius of curvature (m) | | | | | |
|---|---|---|---|---|---|---|
| | Numerical Results | Analytical Solution | Relative Difference (%) | Numerical Results | Analytical Solution | Relative Difference (%) |
| | $h_1$=100 µm, $h_2$=25 µm | | | $h_1$= 40 µm, $h_2$= 10 µm | | |
| 56.85 | 9.54 | 9.46 | 0.88 | 3.79 | 3.78 | 0.19 |
| 106.85 | 5.15 | 5.03 | 2.35 | 2.05 | 2.01 | 1.86 |
| 156.85 | 3.55 | 3.43 | 3.59 | 1.43 | 1.37 | 4.30 |

**2. Detailed description of the total hemispherical emissivity measurements and apparatus:** We employed a comparative calorimetric method under vacuum to measure the effective emissivity of the smart surfaces higher than the ambient temperature. We placed a black matt surface ($\varepsilon_{black}$ = 0.95) on the aluminium thermal pad, which was then placed into the vacuum chamber; we measured the steady state temperature ($T_{black}$ = $T_i$) as a function of the applied power ($Q_{black}^{electrical}\big|_T$) at a certain medium vacuum level (12 ± 0.2 Pa) and ambient temperature ($T_\infty$ ≈ 19.8 °C). The vacuum and the outer temperature remained constant for 200 min prior to starting each experiment. Each temperature level ($T_i$ = 30 °C, 40 °C, 50 °C, 60 °C, 70 °C, 80 °C, 90 °C, and 100 °C) was reached; the steady-state was maintained for at least 30 min to ensure that there would be no temperature change versus time (T = $T_{Steady-state}$ ± 0.01 °C). While maintaining all aforementioned parameters constant, we placed the smart patterned surface on the aluminium thermal pad inside the vacuum chamber; we then identified the required applied power ($Q_{smart}^{electrical}\big|_T$) for each aforementioned temperature level ($T_{smart}$ = $T_i$). The effective emissivity was determined as a function of temperature through the following equation.

$$\left.\begin{array}{l}Q_{black}^{electrical}\big|_T = Q_{black}^{rad} + \left(Q_{silver}^{rad} + Q_{black}^{losses}\right) \\ Q_{smart}^{electrical}\big|_T = Q_{rad}^{smart} + \left(Q_{silver}^{rad} + Q_{black}^{losses}\right)\end{array}\right\} \xRightarrow{T_{smart}=T_{black}=T} Q_{black}^{electrical}\big|_T - Q_{black}^{rad} = Q_{smart}^{electrical}\big|_T - Q_{smart}^{rad} \Rightarrow$$

$$\Rightarrow Q_{smart}^{rad} = Q_{smart}^{electrical}\big|_T - Q_{black}^{electrical}\big|_T + Q_{black}^{rad} \Rightarrow \varepsilon_{eff}^{smart} = \frac{VI^{smart} - VI^{black} + \varepsilon_{black}\sigma A\left(T^4 - T_\infty^4\right)_{black}}{\sigma A\left(T^4 - T_\infty^4\right)_{smart}} \quad (S2)$$

We recorded and compared the measurements of (T, $T_\infty$, V, I) for each effective emissivity value with those of the black coating. It is highly important that $\left(T^4 - T_\infty^4\right)_{black} \approx \left(T^4 - T_\infty^4\right)_{smart}$ between the two comparative measurements. Additionally, we recorded and compared the resistance of the overall circuit with the previous measurements to avoid uncertainties and potential internal faults. This method is very sensitive in temperature levels near the ambient temperature owing to the 4$^{th}$ power of the temperature. Small deviations may result to large discrepancies.

**Thermal pad and measurement acquisition:** We used a flexible silicon heater pad (3 Watts) to control the temperature of the smart surfaces (Fig. S6B). We connected the silicon heater pad using two highly conductive copper wires (6 mm$^2$) with a direct current (DC) voltage power supply. Voltage was applied on the thermal pad owing to the Joule effect, resulting in a uniform temperature increase on the surface. We attached the upper side of the heater pad to an aluminium flat plate of a 1.5 mm thickness. The aluminium sheet was covered with four layers of black matt paint, with an emissivity of ≈ 0.95. The opposite side of the pad was covered with a grooved aluminium plate. We attached a pure highly polished silver sheet with a thickness of 0.3 mm at the lower and external surface of the aluminium thermal pad to prevent heat transfer through radiation as much as possible. Two thin K-thermocouples were placed inside the 1$^{st}$ aluminium sheet. The supplied electrical power must be accurately controlled as a function of the steady-state temperature of each sample. For this reason, we used a stabilised DC power supply (TTi QPX1200SP Bench Power Supply) to accurately regulate the voltage (accuracy Voltage– resolution Voltage - 1 mV). We measured the applied voltage with a multimeter (GW Instec,

GDM-8251A), with an accuracy of ±(0.012% rdg + 5 digits). The current was measured with accuracy of ±100 μA.

**Temperature acquisition and thermocouples:** We placed two K-thermocouples, specially designed for vacuum conditions, inside the thermal pad. We placed the tip of the thermocouple 0.1 mm below the middle of the outer surface of the material to measure the temperature near the radiative surface, and to minimise measurement uncertainties caused by the temperature gradient through the thickness. The junctions of the thermocouples were covered with an ultrathin electrical epoxy-based insulating film. The vacuum chamber operated near room temperature; we continuously recorded the temperature of the inner surface ($T_\infty$) using an array of six thermocouples. The thermocouple arrays measured the temperature of the internal wall of the vacuum chamber (0.1 mm above the internal black surface). The measurements were continuously recorded with a 500 ms sampling rate, and stored in a PC via a calibrated USB data acquisition hardware (Picolog TC-08). All measurements were obtained once steady-state conditions were reached.

**Vacuum Chamber:** To avoid measurement discrepancies, we used a steel vacuum chamber; we coated all inner wall surfaces with three layers of black matt paint, with an emissivity of approximately ε ≈ 0.95, to create a large black-body cavity, and to ensure that the reflections would be negligible (Fig. S6A). We realised this black-body effect by employing a highly absorbing surface, by making the surface area considerably larger than that of the specimen, and by avoiding any external energy radiation sources. The diameter of the chamber was (D = 0.36 m) and the weight was 40 kg. The thermal pad and the investigated materials were located near the centre of the vacuum chamber. The relationship between the vacuum chamber size and its required surface emittance was estimated from the following equation for the shape factor of a grey body, for a surface completely enclosed by another surface. To ensure that the external environment acts as a black body, the following condition must apply (48).

$$\frac{1}{\varepsilon_{material}} \gg \frac{A_1}{A_2}\left(\frac{1}{\varepsilon_{chamber}} - 1\right) \quad (S3)$$

This condition can be satisfied for all possible values of specimen emittance by using an apparatus design in which $A_1/A_2$ = (area of pad)/(internal area of vacuum chamber). The patterned surface may change its emissivity from very low to very high emissivity. The lower surface of the thermal pad was covered with a pure and highly polished silver foil (thickness of 0.3 mm). Assuming that the upper surface of the patterned surface has a maximum total hemispherical emissivity of 0.9, and that the lower surface of the pad has an emissivity value of less than 0.05, we may expect that the vacuum chamber approximately behaves as a large black body, Eq. 16. For the open and closed mode of the patterned surface, the calculated ratios are the following.

$$\frac{1/0.9}{\frac{A_{pad}}{A_{chamber}}\left(\frac{1}{0.95}-1\right)} = 2.5 \times 10^3 \quad and \quad \frac{1/0.075}{\frac{A_{pad}}{A_{chamber}}\left(\frac{1}{0.95}-1\right)} = 30.4 \times 10^3$$

The lower surface of the thermal pad (silver side of the thermal pad) has a ratio equal to the following:

$$\frac{1/0.035}{\frac{A_{pad}}{A_{chamber}}\left(\frac{1}{0.95}-1\right)} = 65.2 \times 10^3$$

For all cases, the calculated ratios are very high, thus ensuring that the chamber behaves as a black body. Vacuum was achieved by using two oil vacuum pumps (2-stage, $1.99 \times 10^{-2}$ mbar) with an overall flow rate equal to 284 L/min; the vacuum level was measured with a Supco VG64 digital vacuum gauge.

**Temperature Uniformity:** The temperature uniformity over the heater pad (Fig. S6B) ensures that the entire surface radiates energy uniformly. In this case, we assumed that the temperature was the same for the entire surface, and could be expressed as a simple parameter, namely T =

T(x,y) for every (x, y). Two different comparisons were carried out: **i.** comparison between two thermocouples, and **ii.** temperature uniformity analysis using thermography. The temperature deviation between the two thermocouples during the heating at the lower and higher temperature level is presented in Fig. S6C. The relative difference between the two temperature measurements during the heating stage or at the steady state was less than 0.43% (Fig. S6C). In both setups, an infrared (IR) camera was located above the samples, and recorded the emitted radiation during heating. The IR camera was FLIR SC660, which has a high resolution pixel detector of 640 × 480 pixels, and has a thermal sensitivity of ≤ 45 mK. The thermal camera images prove that the thermal pad was uniformly heated (Fig. S6D). After analysing the area, we calculated that the average temperature was 92.5 °C with a standard deviation of 0.5, whereas the thermocouple indicated a temperature of 92.7 °C. For the second temperature level, the average temperature was 30.4 °C with a standard deviation of 0.1, whereas the thermocouple indicated a temperature of 30.4 °C. These very small discrepancies did not affect the obtained experimental measurements.

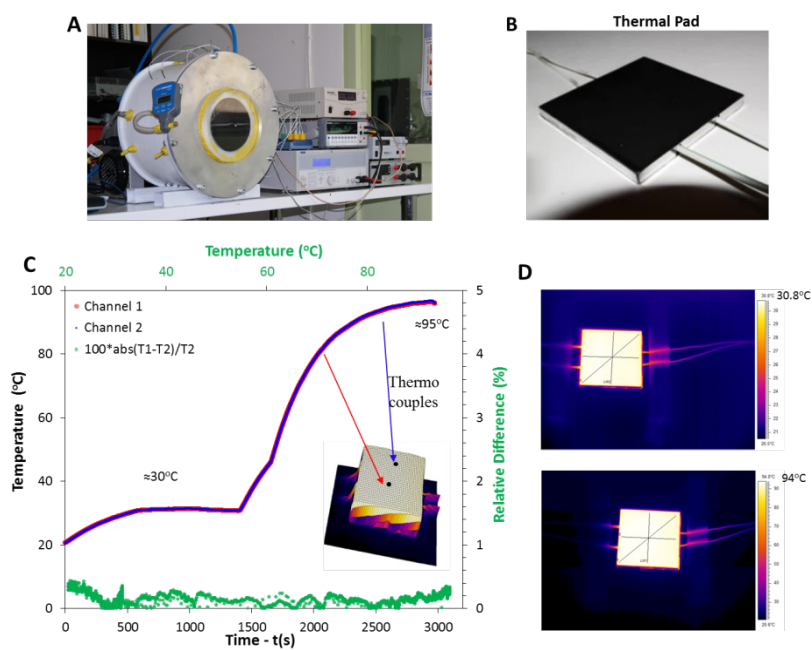

**Fig. S6.** Experimental methodology and apparatus. (A) Vacuum chamber apparatus. (B) Thermal pad device for the activation of the patterned surfaces. (C) Temperature response of the measurement points under random power supply level and relative difference. (D) Temperature field obtained from thermography near the lowest and the highest temperature level.

## II. Supplementary Videos

**Supplementary Movie S1.** Developed pattern surface on a hexagonal lattice, C[112] from high to low emissivity (open to closed state), and thermal cycles under infrared light.

https://www.researchgate.net/publication/317597946_Supplementary_Video_1_-_Developed_pattern_surface_on_a_hexagonal_lattice_C112_from_high_to_low_emissivity_open_to_closed_state_and_thermal_cycles_under_infrared_light

**Supplementary Movie S2.** Developed pattern surface with low to high emissivity and thermographic images during heating (the surface emit more in higher temperature).

https://www.researchgate.net/publication/317598036_Supplementary_Video_2_-_Developed_pattern_surface_with_low_to_high_emissivity_and_thermographic_images_during_heating_the_surface_emit_more_in_higher_temperature

**Supplementary Movie S3.** Developed pattern surface with high to low emissivity and thermographic images during heating (the surface emit less in higher temperature). Despite the fact that all surfaces are at a similar temperature, the thermal camera captures a temperature field that it is not 'correct' owing to the very low emissivity of the outer material of the surface (thermal camera set to $\varepsilon \approx 0.95$).

https://www.researchgate.net/publication/317598118_Supplementary_Video_3_-_Developed_pattern_surface_with_high_to_low_emissivity_and_thermographic_images_during_heating_the_surface_emit_less_in_higher_temperature

**Supplementary Movie S4.** Shape transformation of the multilayer material during heating and cooling. The oriented polyethylene and the aluminium strips present anisotropic thermo-mechanical properties.

https://www.researchgate.net/publication/317598215_Supplementary_Video_4_-_Shape_transformation_of_the_multilayer_material_during_heating_and_cooling

**Supplementary Movie S5.** Shape transformation of the developed surfaces, incorporating rectangular motifs on a strip with dimensions of (54 × 54 mm), and thermographic images. (A) Pattern $G_{[112]}^{[1111]}$ (p111). (B) Pattern $G_{[112]}^{[1122]}$ (pm11). (C) Pattern $G_{[112]}^{[1212]}$ (pm11).

https://www.researchgate.net/publication/317598122_Supplementary_Video_5_-_Shape_transformation_of_the_developed_surfaces_incorporating_rectangular_motifs_on_a_stri

p_with_dimensions_of_54_54_mm_and_thermographic_images_A_Pattern_p111_B_Pattern_pm11_C_

**Supplementary Movie S6.** Transient thermo-mechanical models for pattern $G_{[112]}^{[1111]}$ with fully deformable motifs (8 mm), and calculation of the ambient view factor during the heating stage. Blue line depicts the external surface (position: p3), red line depicts the internal lower surface (position: p1), green line depicts the internal middle surface (position: p2).

https://www.researchgate.net/publication/317598120_Supplementary_Video_6_-_Transient_thermo-mechanical_models_for_pattern_with_fully_deformable_motifs_8_mm_and_calculation_of_the_ambient_view_factor_during_the_heating_stage

**Supplementary Movie S7.** Transient thermo-mechanical models for pattern $G_{[112]}^{[1111]}$ with partially deformable motifs (2 mm), and calculation of the ambient view factor during the heating stage. Blue line depicts the external surface (position: p3), red line depicts the internal lower surface (position: p1), green line depicts the internal middle surface (position: p2).

https://www.researchgate.net/publication/317598119_Supplementary_Video_7_-_Transient_thermo-mechanical_models_for_pattern_with_partially_deformable_motifs_2_mm_and_calculation_of_the_ambient_view_factor_during_the_heating_stage

**Supplementary References**